\documentclass[twocolumn,showpacs,preprintnumbers,amsmath,amssymb]{revtex4}

\begin{document}


\title{
Anomalous heat conduction in one dimensional momentum-conserving systems.
}

\author{Onuttom Narayan${}^{1,2}$ and Sriram Ramaswamy${}^{1}$}
\affiliation{${}^1$Centre for Condensed Matter Theory, Department of Physics, 
Indian Institute of Science, Bangalore 560012, INDIA\\
${}^2$ Department of Physics, University of California, Santa Cruz, CA 95064.${}^*$}

\date{\today}


\begin{abstract}
We show that for one dimensional systems with momentum conservation, the 
thermal conductivity $\kappa$ generically diverges with system size as 
$L^{1/3}.$
\end{abstract}

\pacs{PACS numbers: 05.10.Ln, 75.40.Mg}

\maketitle
When a very small temperature difference is applied across a system,
it is expected that in steady state the heat current $j$ will obey
Fourier's law of conduction
\begin{equation}
j = - \kappa \nabla T
\label{fourier}
\end{equation}
where $T$ is the local temperature and $\kappa$ is the heat conductivity
of the material. Although $\kappa$ is in general temperature dependent,
if the applied temperature difference is small it should be constant
across the system.  Thus if $T_1$ and $T_2$ are the temperatures at which
the two ends of a system of length $L$ are kept (with $T_1 \geq T_2$),
the steady state current should be $j = \kappa (T_1 - T_2)/L.$

On the other hand, for many one-dimensional models it can be shown
analytically~\cite{Exact,Dhar,n4} or numerically~\cite{n4,n1,Dhar2,n2,n3}
that $j\propto L^{\alpha - 1}$ with $\alpha> 0,$ even in the linear
response regime where $j\propto T_1 - T_2.$ The value of $\alpha$
differs from model to model. In the framework of Eq.(\ref{fourier}),
this would imply an $L$-dependent conductivity that diverges in the
infinite system limit.  (In some oscillator models, $\alpha<0,$ implying
an anomalous but not divergent conductivity.)

Recently, it has been argued~\cite{Prozen} that such anomalous heat 
conduction occurs only in systems with momentum conservation, and is
a consequence thereof. This was done by showing that if $j(x, t)$ is
the energy current density, the autocorrelation function of the total
energy current $J(t) = \int dx\, j(x, t)$ 
\begin{equation}
C(t_1 - t_2) = \langle J(t_1) J(t_2) \rangle
\label{corfn}
\end{equation}
has the property $C(t\rightarrow\infty)\neq 0.$ Although the proof of this
result in Ref.~\cite{Prozen} was specific to one dimensional systems,
Galilean invariance allows one to construct a general proof for any
dimension $d$~\cite{Green}.  This is because the energy current $J(t)$
has an advective contribution $(E + p L^d) \overline v,$ where $E$
is the energy, $p$ the pressure, $L^d$ the volume and $\overline v$
the center of mass velocity of the system. For an energy and momentum
conserving system, this advective contribution to $J(t)$ is time
independent. $C(\infty)$ is found by calculating $\langle [(E + p L^d)
\overline v]^2\rangle$ within the canonical ensemble, and is non-zero.
The limit $\lim_{\tau\rightarrow\infty}\lim_{L\rightarrow\infty} (1/(T^2
L))\int^{\tau}_0 dt C(t)$ is then divergent.  The Kubo formula~\cite{Kubo}
was invoked~\cite{Prozen} to equate this to $\kappa.$

Although (as discussed in the next paragraph) the argument in
Ref.~\cite{Prozen} is incorrect and a nonvanishing $C(\infty)$ 
has no consequence for heat conduction, the conclusion
that momentum conservation in low dimensional systems (generically)
implies anomalous conduction is valid. This is because, in addition to
the limit $C(\infty)$ being finite, there is also a slowly decaying
tail in $C_0(t) = C(t) - C(\infty).$ Unlike $C(\infty)\neq 0,$ which
is valid in all dimensions, the tail in $C_0(t)$ decays sufficiently
slowly to cause a singularity in $\kappa$ only for $d \leq 2.$ In this
paper, using the transport equations for a normal fluid with thermal
noise added, we show that $\kappa(L)\propto L^\alpha,$ with $\alpha =
(2 - d)/(2 + d) > 0$ for $d < 2.$ (There is a logarithmic singularity
for $d=2.$) For the physical case of $d=1,$ we obtain $\alpha = 1/3.$
These transport equations should be valid for all systems which rapidly
reach local thermal equilibrium, and for which the only slowly evolving
quantities are the mass energy and momentum densities.  The results are
therefore generic, although the assumption of local thermal equilibrium
breaks down for some models, such as hard sphere equal mass particles
in one dimension, or a chain of harmonic oscillators, for which $\alpha$
is different~\cite{Exact}.

We first recall the discussion of Bonetto {\it et al\/}~\cite{Bonetto}
about the argument in Ref.~\cite{Prozen}. The conductivity in momentum
conserving systems is in fact not obtained from the autocorrelation
function of  $J(t),$ but of $J_0(t)=J(t) - (E + pL^d) \overline
v.$~\cite{foot1} This result can be proved rigorously within linear
response theory~\cite{statbook}.  If one makes a linearized hydrodynamic
approximation, which is equivalent to linear response theory with no
(thermal) noise terms in the transport equations, the derivation of this
result is simpler and well known~\cite{kadanoff,lutt}.  When $J(t)$
is replaced by $J_0(t)$, the resultant truncated correlation function
$C_0(t)$ has no infinite time tail, and
\begin{equation}
\kappa = \lim_{\tau\rightarrow\infty}\lim_{L\rightarrow\infty}
{1\over{T^2 L}}\int^{\tau}_0 dt C_0(t)
\label{cured}
\end{equation}
is cured of its divergence.

Despite the fact that $C_0(\infty) = 0,$  we show in this paper that, with
thermal noise in the transport equations, $C_0(t)$ decays sufficiently
slowly with $t$ to make the integral $\int dt\, C_0(t)$ --- and therefore
the conductivity $\kappa$ --- divergent for $d \leq 2.$ We
first show this qualitatively. The advective contribution to $J(t)$
is really equal to $\int d^d x h(x) v(x),$ where $h(x)$ is the local
enthalpy density and $v(x)$ the local velocity. (Vector indices have
been suppressed.)  This can be expressed as $\int d^d k h(k) v(-k) +
(E + p L^d) \overline v,$ where the first integral is restricted to
$k\neq 0,$ and the second part comes from the non-zero spatial average
of $h(x).$ In going from $J(t)$ to $J_0(t),$ only the second part of this
was removed.  Since the time decay of all the
hydrodynamic modes is diffusive~\cite{Forster}, expressing $h(k, t)$ and
$v(k, t)$ in terms of hydrodynamic modes and approximating $\langle h(k,
t) v(-k, t) h(k^\prime, 0) v(-k^\prime, 0)\rangle$ as $\langle h(k, t)
h(k^\prime, 0) \rangle \langle v(-k, t)v(-k^\prime, 0)\rangle$ yields
the advective contribution to $C_0(t)$ to be $\sim L^d \int d^d k\,
\exp[-O(k^2) t],$ which is $\sim L^d/t^{d/2}.$ From Eq.(\ref{cured}),
$\kappa$ diverges for $d \leq 2$~\cite{foot3}.

Although the conductivity $\kappa$ does indeed diverge for $d\leq 2$
as this rough calculation indicates, the tail of $C_0(t)$ decays as
$t^{-2d/(d + 2)}$ instead of as $t^{-d/2}.$ This is because thermal
noise in the transport equations gives rise to singular corrections to
the parameters in the equations, so that the hydrodynamic modes decay
superdiffusively. As is standard in renormalization group (RG) analyses
of such phenomena, we solve the linearized transport equations and check
whether the non-linear corrections are relevant for long wavelength
low frequency phenomena. Below $d=2,$ which is thus the upper critical
dimension, the nonlinearities are found to be relevant, and their effect
is calculated.

We assume that the system whose thermal conductivity is of interest
is one which reaches local thermal equilibrium, with the only dynamical
variables that evolve slowly with time being the mass, energy and momentum
densities. With these assumptions, the appropriate transport equations
for the system are those for a normal fluid~\cite{Forster}, with thermal
fluctuations included in the form of noise sources~\cite{landl,pep}:
\begin{eqnarray}
\partial_t \rho + \nabla\cdot (\rho v) &=& 0\nonumber\\
\partial_t (\rho v) + \nabla\cdot(\rho v v)
& = &  -\nabla p +  (\zeta + \eta/3) \nabla\nabla\cdot v \nonumber\\
&+& \eta \nabla^2 v + \zeta_v\nonumber\\
\partial_t\epsilon + \nabla\cdot[(\epsilon +p) v] &=& 
\nabla\cdot\kappa_0\nabla T  + O((\nabla v)^2) + \zeta_\epsilon
\label{lutteqs}
\end{eqnarray}
where $\rho$ is the local density of the fluid. The local temperature $T$
and pressure $p$ are implicit functions of $\rho, v$ and $\epsilon.$ The
thermal noise terms $\zeta_{v,\epsilon}$ satisfy $\langle \zeta_v(x_1,
t_1) \zeta_v(x_2, t_2) \rangle\propto -k_B T \delta(t_1 - t_2)
\partial^2 \delta(x_1 - x_2),$ and similarly for $\zeta_\epsilon,$
with the proportionality constants fixed by the requirement that
the variance of the fluctuations $\delta\rho =\rho - \rho_0,$
$\delta\epsilon=\epsilon - \epsilon_0$ and $v$ at any instant are those
of a system in equilibrium at temperature $T.$ ($\nabla\cdot(\rho v v)$
is a vector whose $i$'th component is $\partial_j(\rho v_i v_j).$ The
first equation in Eqs.(\ref{lutteqs}) is an exact identity, and has no
thermal noise correction.

It is standard and straightforward to solve these equations with
the linear approximation, valid if $\delta\rho(x, t),$ $v(x,t)$ and
$\delta\epsilon(x, t)$ are small. One obtains~\cite{kadanoff,Forster} two
propagating sound modes and one non-propagating heat mode. All three modes
decay diffusively.  The relevance or irrelevance of the nonlinear terms
in the transport equations is then determined by calculating corrections
to correlation and response functions to one loop in a diagrammatic
perturbation expansion. For instance, the response of $v(x)$ to a
perturbation in the second member of Eqs.(\ref{lutteqs}), $g_{vv}(x_1 -
x_2; t_1 - t_2),$ receives a one loop correction from $\nabla\cdot(\rho v
v),$ resulting in corrections to the viscosities $\eta,\zeta$ of the form
\begin{equation}
\delta\eta,\delta\zeta\sim \int d^d x d t c_{vv}(x; t)
g_{vv}(x; t)
\end{equation}
where $c_{vv}$ is the autocorrelation function of $v(x).$ (The component
indices in $c_{v_i, v_j}$ and $g_{v_i, v_j}$ have been suppressed for
compactness.) Expanding the correlation function $c$ and the response
function $g$ in terms of the three hydrodynamic modes, and performing the
$x$ integral first, the integrand is negligible outside a region of volume
$O(t)^{d/2}.$ (The propagating parts of the modes shift the peak of the
integrand away from $x=0$ if the contribution of the same hydrodynamic
mode is considered for $c$ and $g.$ This is inconsequential if the
system is large.) Since the correlation and response functions of all the
hydrodynamic modes have a $|t|^{-d/2}$ prefactor, the integral over $x$
yields $\delta\eta,\delta\zeta\sim \int dt \theta(t) t^{-d/2},$ where the
$\theta$-function comes from causality in the response function. The $t$
integral diverges for $d \leq 2.$ Similar calculations can be carried out
for other one loop corrections.  Since all the autocorrelation functions
have the same $|t|^{-d/2}$ prefactor, the RG scaling dimension of all
the three density fields is $-d/2:$  $|t|^{-d/2}\sim (|x|^{-d/2})^2.$

For $d \leq 2,$ the nonlinear corrections are therefore relevant when
expanding around the linearized equations. This can also be seen by
scaling all the variables in the transport equations as $x = \lambda
x^\prime,$ $t = \lambda^z t^\prime,$ $\zeta_{v,\epsilon} = \lambda^{-(d+2
+ z)/2} \zeta^\prime_{v,\epsilon},$ and $(\delta\rho, v, \delta\epsilon)
= \lambda^{-d/2} (\delta\rho, v, \delta\epsilon).$ The time derivative,
dissipative and thermal noise terms in Eqs.(\ref{lutteqs}) scale
identically if the dynamic exponent $z$ is set to 2. (The terms with
one spatial derivative in the linearized equations grow, since they
control the propagation of the sound modes, whose speed is obviously
altered if $t$ is scaled as $x^2.$ However, as we have seen in the
previous paragraph, this does not affect the scaling of the loop
corrections.) The nonlinear terms can be seen to be relevant if $d < 2.$

For $d<2,$ a renormalization group analysis has to be carried out,
integrating out loop corrections from short wavelength fluctuations along
with rescaling the variables.  The nonlinearities grow under the RG flows
until they reach a non-trivial fixed point. The new scaling dimensions
of the fields and the dynamic exponent $z$ can then be evaluated at this
fixed point. It is, however, not necessary to carry out such a calculation
to obtain the exponents: they can be determined completely from symmetry
considerations. The RG flows preserve the property that equal time
fluctuations in $\delta \rho, v $ and $\delta\epsilon$ must be those
of a system in equilibrium at temperature $T.$ Since the fluctuations
in these densities must be Gaussian at sufficiently long wavelength,
we require that $\int d^d x [v^2, (\delta\rho)^2, (\delta\epsilon)^2]$
should be invariant under rescaling.  Thus the scaling dimensions of
all three fields are equal to $-d/2$ even for $d<2.$ Further, Galilean
invariance relates the loop corrections to $\partial_t \phi$ and to the
corresponding advective term $\nabla\cdot(v\phi)$ for any (conserved)
field $\phi$~\cite{FNS} Since both terms are invariant at the fixed point,
this yields the condition that $\nabla v $ scales as $\partial_t,$ i.e.
$v$ scales as $x/t.$ Combining this condition with the previous one,
we obtain $z = 1 + d/2.$

The energy current density is obtained by requiring that the third
equation in Eq.(\ref{lutteqs}) should be equivalent to $\partial_t
\epsilon + \nabla\cdot j = 0.$ This yields $j(x, t) = (\epsilon + p)
v - \kappa_0\nabla T + O(\nabla v^2, v \nabla\cdot v).$ Since $J_0(t)
= J(t) - (E + p L^d) \overline v,$ we obtain correspondingly $j_0(x,
t) = (\delta\epsilon + p - p_0) v - \kappa_0\nabla T + O(\nabla v^2,
v \nabla\cdot v).$ Under the RG rescaling, the three terms in this
scale with dimension $-d, -1 - d/2$ and $ -1 -d $ respectively, and the
first term is most important for $d < 2$.  If one expresses the transport
equations Eqs.(\ref{lutteqs}) through a generating functional ~\cite{MSR},
and adds an extra $a (\delta\epsilon + \delta p) v$ term in the argument
of the exponent, the $v\rightarrow - v, x\rightarrow -x$ symmetry of
Eqs.(\ref{lutteqs}) ensures that this term is not renormalized to $O(a).$
Thus the scaling of $C_0(t)/L^d $ can be obtained from the bare scaling
dimension of $j_0(x, t):$
\begin{equation}
C_0(t)/L^d =\int d^d x \langle j_0(x, t) j_0(0, 0)\rangle\sim |t|^{\alpha - 1}
\label{coreqn}
\end{equation}
with
\begin{equation}
\alpha = 1 - d/z = (2 - d)/(2 + d).
\end{equation}

Eq.(\ref{coreqn}) has to be integrated over $t$ to obtain $\kappa$ from
Eq.(\ref{cured}).  For a system of linear dimension $L,$ the tail of
$C_0(t)$ obtained in Eq.(\ref{coreqn}) is valid only when a disturbance
at $x=0, t=0$ in the propagating modes has not been carried outside the
system. This is because the tail in $\langle h(x, t) v(x, t) h(0, 0)
v(0, 0)\rangle$ comes from long wavelength fluctuations, for which the
contribution to $v$ from the heat diffusion mode is zero, so that $v$
must be expressed as a linear combination of the two propagating modes. A
fluctuation in $v$ is carried to the boundaries of the system in a time
$t\sim O(L).$  The precise behavior thereafter depends on the coupling
to the heat baths at the boundaries, but in any case the fluctuation
is partially or fully lost to the baths.  The tail of $C_0(t)$ is cut
off in a few round trip times, i.e in a time $t\sim O(L).$ Substituting
Eq.(\ref{coreqn}) in Eq.(\ref{cured}) and using this cutoff, we obtain
\begin{equation}
\kappa(L) = L^\alpha.
\end{equation}
Thus the conductivity measured in a system of size $L$ diverges with
$L,$ or equivalently, the heat current flowing across a system with 
a fixed small temperature difference decays as $\sim L^{\alpha - 1}$ with
$L.$ For the physically relevant case of $d = 1, $ the heat current 
must decay as $\sim L^{-2/3}.$ For $d=2,$ the conductivity $\kappa$ has
a logarithmic singularity as a function of $L.$

We now compare with earlier analytical and numerical results. For
a chain of coupled harmonic oscillators, there has been a large
amount of analytical work showing that $\kappa$ diverges with system
size~\cite{Exact}. The form of the divergence is different for different
models, and in fact varies over a wide range depending on the heat
baths~\cite{Dhar}. However, all these are systems where local thermal
equilibrium is not established, so the results of this paper do not
apply. With more complicated models, there have been various numerical
simulations that have shown a divergent $\kappa,$ with $\alpha$ ranging
from 0.17 to 0.5~\cite{n4,n1,Dhar2}. However, the simulations in these
papers were performed for fairly small system sizes, up to $\sim 1000.$ The
scaling of $\kappa$ as a function of $L$ is not very good, indicating the
need for larger system sizes. Very recently, there have been simulations
on a one dimensional hard sphere gas with alternating masses~\cite{f6}
for much larger system sizes: up to 16383~\cite{n3} and 30000~\cite{n2}. In
the former, the scaling is not very good, and the dependence of $\kappa$
on $L$ varies considerably with the mass ratio between neighboring
particles. However, the authors estimate $\alpha$ to be 0.31 to 0.35. In
the latter paper, the scaling is very good, and $\alpha$ is 
0.255, which disagrees with our paper. It is surprising that
numerics on the same model, with roughly the same range of sizes, yields
such different results. The current $j(x, t)$ is chosen differently in
both papers, and it is not clear how it is defined in the latter paper,
since it involves derivatives of the (singular) hard sphere potential.
Further work is needed to clarify the situation.

In Ref.~\cite{n4}, there is a mode-coupling calculation that relies on
Ref.~\cite{ernst}, indicating that $\alpha$ should be 2/5.  The argument
in \cite{ernst} is internally inconsistent: the scaling $\omega\sim
D(\omega) k^2$ is used to correctly find the renormalized diffusion
coefficient $D(\omega)\sim \omega^{-1/3}$ and thereby the long-time
tail of $C_0(t)$ as $\sim t^{-2/3},$ but then $\omega\sim k$ is used to
incorrectly convert $D(\omega)$ to $D(k)\sim k^{-1/3}.$ This expression
is then used by Ref.~\cite{n4} to obtain $C_0(t)\sim t^{-3/5}.$ As
we have seen, although $t\sim x$ {\it is\/} the scaling conversion
to be used in Eq.(\ref{cured}), this is not appropriate for the loop
integrals or the dynamic exponent.  Secondly, although their system
is nominally a 1-dimensional {\em crystal}, it should behave like a
fluid at large length scales since fluctuations will wipe put long-range
order.  Nonetheless, if we treat it as a crystal, we will encounter the
``Poisson-bracket'' nonlinearity $\nabla u \delta H / \delta u$ in the
equation for the momentum density, where $H$ is the elastic Hamiltonian
for the displacement field $u$. This term is nonlinear even if we retain
only harmonic terms in $H$. Simple power-counting shows that
this yields nonlinear corrections of precisely the same type as those
from the advective term. In fact, the analysis of Ref.~\cite{n4} is
equivalent to this.

For a system in which momentum is not conserved, there is no advective
term in the energy or mass current both of which depend on gradients
of $\rho$ and $\epsilon.$ Even the nonlinear terms in the equations
for $\partial_t\rho$ and $\partial_t\epsilon$ thus have at least
two spatial derivatives, and are irrelevant compared to the linear
terms. The conductivity is given by Eq.(\ref{cured}) with $C(t)$
instead of $C_0(t)$~\cite{statbook,lutt} This neither has a non-zero
$C(\infty)$ limit, nor a slowly decaying tail for large $t.$ Indeed,
numerical studies~\cite{noncons} confirm that for such systems, the
conductivity is finite.

If a system is integrable, even if it does not conserve momentum, the
conductivity can be singular, since Refs.\cite{statbook,lutt} assume
local thermal equilibrium.  Also, recent numerical studies on a momentum
conserving chain of coupled rotators~\cite{rotator} show no anomalous
conductivity at high temperatures.  The interparticle potential is
$V(q_{i+1} - q_{i}) = 1 - \cos(q_{i+1} - q_{i}),$ and the particles at
the end are kept in contact with different heat baths (through Langevin
noise). Physically, the $q$'s can be interpreted as angles.  Although the
model is momentum conserving, since the heat baths are applied to specific
particles rather than at the ends, there is no advective term in the
current: a long wavelength momentum fluctuation causes the ``angles"
to ``spin" round and round instead of carrying energy from one side
to another.  At low temperatures, when fluctuations are small and the
model goes over to a harmonic oscillator, a long wavelength momentum
fluctuation is transmitted to neighboring particles and thence to
the ends of the chain, effectively leading to advective transport. The
nature of the transition between the high and low temperature phases is
interesting and requires further work.  Also, if a system is integrable,
even if it does not conserve momentum, the conductivity can be singular,
since Refs.\cite{statbook,lutt} assume local thermal equilibrium.

In this paper, we have shown that for one-dimensional momentum conserving
systems, the heat current when a small temperature difference $\delta
T$ is applied across a system of length $L$ is generically of the form
$j\propto (\delta T)/L^{2/3}.$ This is consistent with earlier
numerical studies, but further work is needed to improve the numerical
picture.

We thank Abhishek Dhar and Sriram Shastry for very useful comments and 
discussions.


\begin{references}
\bibitem{foot1} Alternatively, this can be written as the autocorrelation 
function of $J(t)$ restricted to the sector of the canonical ensemble 
in which $w$ is zero.
\bibitem{foot3} The propagating parts of the hydrodynamic modes do not
affect this argument when the contribution of the same mode to $\langle
hh\rangle$ and $\langle vv\rangle$ is considered, as can be seen by
going to a moving frame. When different hydrodynamic modes are taken for
$\langle hh\rangle$ and $\langle vv\rangle,$ the time dependence of the
decay is $\sim t^{-d}.$
\bibitem{f6}One can verify numerically that the single-particle velocity
distribution for a long chain, where the temperature gradient is small,
approaches a Gaussian distribution, indicating that local thermal
equilibrium is established~\cite{Dhar2}. This is not the case when all
particles in the chain have the same mass.
\end{references}
\end{document}